\documentclass[conference]{IEEEtran}
\usepackage{graphicx}
\usepackage{amsfonts,amssymb,amsmath,amsgen,amsopn,amsbsy,theorem,graphicx,epsfig}
\usepackage{multirow}
\usepackage{hyperref}

\ifCLASSINFOpdf
\else
\fi

\usepackage{xcolor}

\begin{document}
%
\title{Residual Learning and Filtering Networks for End-to-End Lossless Video Compression}

\author{
\IEEEauthorblockN{Md Baharul Islam}
\IEEEauthorblockA{Bahcesehir University, Istanbul, Turkey\\
Florida Gulf Coast University, Fort Myers \\ FL 33965, United States}
\and
\IEEEauthorblockN{Afsana Ahsan Jeny}
\IEEEauthorblockA{Bahcesehir University, Istanbul, Turkey \\ University of Connecticut, Storrs\\ CT 06269, United States}}


%


\maketitle

\begin{abstract}
Existing learning-based video compression methods still face challenges related to inaccurate motion estimates and inadequate motion compensation structures. These issues result in compression errors and a suboptimal rate-distortion trade-off. To address these challenges, this work presents an end-to-end video compression method that incorporates several key operations. Specifically, we propose an autoencoder-type network with a residual skip connection to efficiently compress motion information. Additionally, we design motion vector and residual frame filtering networks to mitigate compression errors in the video compression system. To improve the effectiveness of the motion compensation network, we utilize powerful nonlinear transforms, such as the Parametric Rectified Linear Unit (PReLU), to delve deeper into the motion compensation architecture. Furthermore, a buffer is introduced to fine-tune the previous reference frames, thereby enhancing the reconstructed frame quality. These modules are combined with a carefully designed loss function that assesses the trade-off and enhances the overall video quality of the decoded output. Experimental results showcase the competitive performance of our method on various datasets, including HEVC (sequences B, C, and D), UVG, VTL, and MCL-JCV. The proposed approach tackles the challenges of accurate motion estimation and motion compensation in video compression, and the results highlight its competitive performance compared to existing methods.
\end{abstract}


%
\IEEEpeerreviewmaketitle

\section{Introduction}
\label{2}

Raw video content, such as 1080p at 30 Hz using the YUV 420 format, can consume around 93 megabytes (MB) per second, making it challenging to handle in real-time scenarios. As per a report \cite{networking2016cisco}, video content already accounts for over 80\% of internet traffic, and this percentage is expected to increase further. Therefore, it is crucial to develop effective video compression systems that can produce higher-quality frames while operating within specific bandwidth constraints. Additionally, video compression techniques have proven to be beneficial for applications like action identification \cite{wu2018compressed} and model compression \cite{han2015deep}.

Over the past few decades, several well-known video compression methods have been proposed, including H.264 \cite{wiegand2003overview}, H.265 \cite{sullivan2012overview}, and VVC \cite{bross2019versatile}. However, they often introduce undesirable artifacts like blocking, ringing, and blurring effects near boundaries during block-wise operations \cite{lim2011ringing}, leading to degraded video quality and a negative impact on user experience. Moreover, these methods cannot be optimized in an end-to-end manner, which calls for further investigation to improve compression performance. Recent research has focused on deep neural network (DNN) based image compression algorithms \cite{theis2017lossy,jeny2022improving}, offering two main advantages: the ability to utilize non-linear transforms and the elimination of the need for handcrafted features. They also concentrate on developing end-to-end optimized frameworks that improve all modules involved in compression, including transformation, quantization, and entropy estimation. Extending image compression techniques to the video domain raises challenges related to temporal consistency. Developing an end-to-end video compression system faces several challenges. Firstly, optimizing the entire system is challenging. While some work \cite{jeny2022improving, jeny2022efficient, jeny2023optimized} have substituted one or two modules in the conventional architecture, an end-to-end improvement of the compression system is essential. Secondly, the video compression performance depends on the accurate motion vector (MV) information. Optical flow-based methods \cite{jeny2021deeppynet} can compute motion information, but inaccurate optical flow estimation can lead to motion-compensated algorithms introducing additional bits that impact the reconstructed frame. Thirdly, a rate-distortion strategy is significant to reduce temporal redundancy.

\begin{figure*}[tb]
    \centering
    \includegraphics[width=1\textwidth]{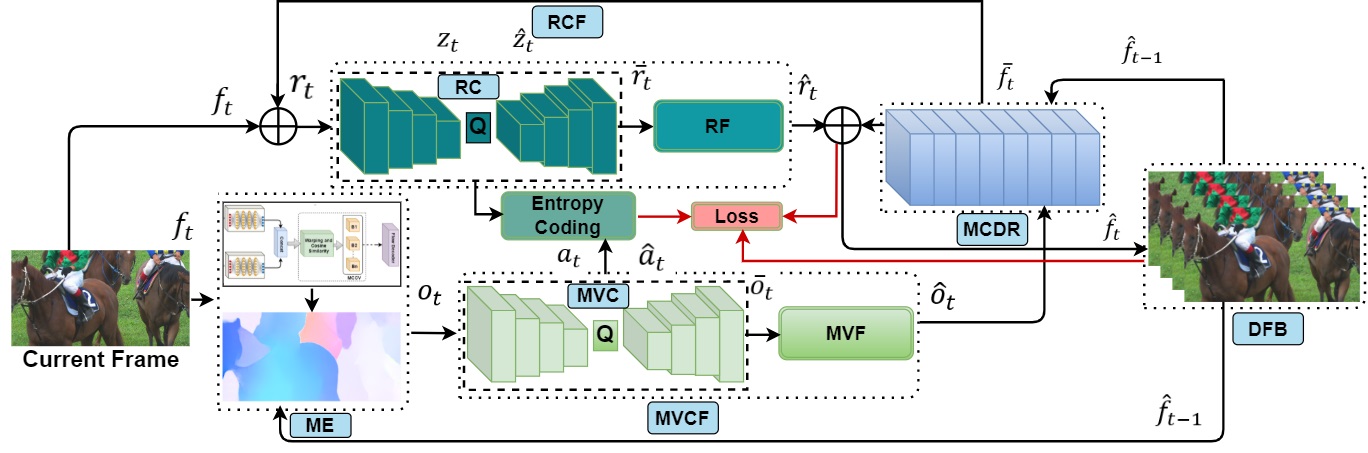}
    \caption{Proposed architecture. In the ME module, we pass the current frame $f_{t}$ and the previous reference frame $\hat{f}_{t-1}$ to the optical flow network to get $o_{t}$. Then it sends to the MVC network to compress the raw optical values. After that $\bar{o}_{t}$ has been fed into the proposed {MVF} to eliminate compression artifacts and received $\hat{o}_{t}$. Next, $\hat{o}_{t}$ is passed through the proposed MCDR network to obtain $\bar{f}_{t}$ and then transmitted to {RCF} module with $r_{t}$. {From that $\bar{r}_{t}$ is acquired from residual compression (RC) network and fed into our proposed method RF to receive $\hat{r}_{t}$}  since it has artifacts due to the quantization of the residual encoder-decoder network. Finally, $\hat{r}_{t}$ and $\bar{f}_{t}$ are added to create the final reconstructed frame, $\hat{f}_{t}$. DFB is utilized as an online buffer to fine-tune the reference frames for making the clear reconstructed frame.}
    \label{fig:f1}
\end{figure*}

{Modern studies \cite{jeny2023optimized, lin2020m} have concentrated on refining the balance between rate and distortion. They often employ deep autoencoder-like structures in both the compression of motion vectors and residuals, drawing inspiration from methodologies in image compression research \cite{balle2016end}. On occasions, they have leveraged transformers. However, the considerable size of these transformer models poses potential challenges when introduced to environments with constrained resources.} However, these deep autoencoder-style networks can be too simple, leading to compression errors in both motion and residual modules, especially at low bit rates after quantization. Therefore, there is a need for a light compression network to enhance the video compression ratio. Additionally, reducing compression errors is crucial for better visualization of the reconstructed frame. To address these challenges, an end-to-end deep rectified video compression network is proposed, aiming to minimize spatial-temporal redundancy. The technical contributions of the paper are summarized as follows:

\begin{itemize}
\item Introduction of an end-to-end video compression system: The proposed system incorporates all components, including motion estimation, compression, compensation, residual compression, and entropy coding, based on the rate-distortion trade-off. 

\item Motion compression using an autoencoder-style network: A motion compression network is proposed, employing residual blocks and skip connections to extract high-resolution and high-semantic features from different layers. This leads to improved motion compression performance and faster convergence.

\item Introduction of filtering networks for reducing compression artifacts: Two filtering networks, namely motion vector filtering (MVF) and residual filtering (RF), are proposed to mitigate compression artifacts in the reconstructed frames. These networks improve the visual quality of the reconstructed frames.

\item Utilization of deep rectifiers and multiple residual blocks: Deep rectifiers, such as PReLU, are employed in the motion compensation and filtering networks to improve computational efficiency by minimizing training parameters. Multiple residual blocks are used to make the predicted frame smoother.

\end{itemize}

\section{Proposed Method}
\label{Proposed Method}
Fig. \ref{fig:f1} shows the proposed method that includes motion estimation (ME), MV compression (MVC), and MV filtering (MVF) network in MV compression and filtering (MVCF) module, motion compensation network (MCDR), and residual filtering (RF) network in residual compression and filtering (RCF) module. These components are tuned together with the whole system and utilize a single rate-distortion loss.

\noindent \textbf{Notations.} 
In the given context, let $\mathrm{F}=\left\{f_{1}, f_{2}, \ldots f_{t-1}, f_{t} \right\}$ represent the original video sequences. At time step $t$, $f_{t}$ represents the original frame, $\bar{f}{t}$ represents the predicted frame, and $\hat{f}_{t}$ represents the reconstructed or decoded frame. The difference between $\bar{f}_{t}$ and $f_{t}$ is denoted by $r_{t}$, a residual frame. The decoded residual frame is represented by $\hat{r}_{t}$. Motion information is crucial for eliminating temporal redundancy, and it is obtained through pixel-wise optical flow estimation. The motion vector or optical flow value is denoted by $o_{t}$. The rebuilt motion information and the final reconstructed or decoded values of the motion vector fields at time step $t$ are represented by $\bar{o}_{t}$ and $\hat{o}_{t}$, respectively. Since the autoencoder framework involves transformation, the residual frame $r_{t}$ and the motion vector frame $o_{t}$ are transformed into $z_{t}$ and $a_{t}$, respectively. After quantization in the motion vector and residual autoencoder, $\hat{z}_{t}$ and $\hat{a}_{t}$ represent the quantized forms of $z_{t}$ and $a_{t}$, respectively.

\subsection{Motion Estimation}
A key component of video compression is motion estimation (ME), which collects motion information. Optical flow is often used in computer vision applications to utilize the temporal correlation between video sequences. In this paper, the recent optical flow method DeepPyNet \cite{jeny2021deeppynet} is utilized to achieve optical flow estimation, which fundamentally helps to reduce the rate-distortion trade-off and improve the entire video compression method's performance. 

\subsection{MV Compression and Filtering}
\textbf{MV Compression Network:} After motion estimation, $o_{t}$ is acquired via optical flow. We design an autoencoder-style network (deep encoder and decoder) to compress the optical flow. 
An example is the GDN, which is very effective and gives considerable benefits when combined with scalable quantization for the image compression network \cite{balle2016end}. However, several changes are carried out in our architecture. Firstly, $o_{t}$ is passed through a sequence of convolution operations and nonlinear transforms. Four convolution and deconvolution layers are used in the encoder and decoder, followed by the GDN or inverse GDN (IGDN) except for the last layer. Then, the paradigm of identity shortcut connection in the ResNet \cite{liu2019adaptive} is added to certain GDN and IGDN layers, represented by the ResGDN and ResIGDN, to enable deeper learning with the finer features of the network. Rather than using ReLU and batch normalization in conventional residual blocks, we employ GDN and IGDN layers in our residual blocks, which give superior performance and a quicker convergence rate. 


\textbf{MV Filtering Network: } At the end of the {MVC}, we obtain the rebuilt/reconstructed motion vector frame $\bar{o}_{t}$. Due to quantization, $\bar{o}_{t}$ may have compression errors, particularly at low bit rates. For instance, once we add the optical flow with the video compression network, several zeros are added to $o_{t}$ so that motion information can be encoded with fewer bits. However, we noticed that these zeros had an erroneous effect throughout the motion compensation process. As a result, we propose an MV filtering (MVF) network to decrease compression errors from $\bar{o}_{t}$. 
In the {MVF} network, the $\bar{o}_{t}$ and $\hat{f}_{t-1}$ are used for filtering. The purpose for employing $\hat{f}_{t-1}$ is that the next motion compensation module will rely on the filtered $\hat{o}_{t}$ and $\hat{f}_{t-1}$ to get the predicted frame, therefore, $\hat{f}_{t-1}$ must be used as a guide to improving $\bar{o}_{t}$. In {MVF,} we concatenate $\bar{o}_{t}$ and $\hat{f}_{t-1}$ and then pass it to six convolution layers with four parameters like $\mathrm{k} \times \mathrm{c} \times \mathrm{s} \times \mathrm{d}$, where $k, c, s$, and $d$ represent the size of the kernel, the output channel, the stride and the dilation rate, respectively. Each convolution layer is followed by a Parametric Rectified Linear Unit (PReLU) instead of Rectified Linear Unit (ReLU) to obtain fewer errors. The ablation study in \ref{Ablation Study} shows that using {MVF} with PReLU has enhanced the video compression system's efficiency. More information about PReLU is expressed in the motion compensation section. Finally, we get $\hat{o}_{t}$ after applying the {MVF}.


\subsection{Motion Compensation Network}
In the motion compensation module, $\hat{f}_{t-1}$ is warped to $f_{t}$ depending on $\hat{o}_{t}$ as Eq. \ref{eq1}.
\begin{equation}\label{eq1}
    \hat{f}_{t-1}^{w}=\operatorname{Warp}\left(\hat{f}_{t-1}, \hat{o}_{t}\right)
\end{equation}
Where, $\hat{f}_{t-1}^{w}$ represents the warped frame and $Warp$ is the backward warp operation \cite{jaderberg2015spatial}. 
Though our proposed CNN motion compensation model MCDR is based on this \cite{lu2019dvc}, we have made significant changes to improve the compression efficiency. The residual operation \cite{he2016deep} has already proved its effectiveness. Therefore, three convolution layers and three activation functions make up our proposed residual block in MCDR, with the final activation occurring after the summation. By employing residual blocks, we can simultaneously allow the network to handle high-resolution features from the previous layer and high-semantic features from later layers. 
The Eq. \ref{eq2} represents the residual block.  
\begin{equation}\label{eq2}
    y_{i+1}=\mathrm{a}\left[y_{i}+r\left(y_{i}, u_{i}\right)\right]
\end{equation}
Where the input of the $i$-th residual block is denoted by $y_{i}, u_{i}=\left\{w_{i, p} \mid 1 \leq \mathrm{p} \leq \mathrm{C}\right\}$ is a collection of weights and biases that are associated with the $i^{th}$ residual block, and $\mathrm{C}$ is the convolution layer's number included inside a residual block (here, $\mathrm{C}$ is $3$). The residual function is denoted by $\mathrm{r}$. The activation function is denoted by $a$. The Rectified Linear Unit (ReLU) is a unit of measure widely used in the residual block. However, as described in \cite{he2015delving}, PReLU is more accurate in the training phase. Hence, we employ the PReLU instead of ReLU in our MCDR network to improve compression efficiency. 
The PReLU is defined below. 
\begin{equation}\label{eq3}
    a (I)= max (I,0)+ S \times {min} (0,I)
\end{equation}
$I$ is the function a's input, while S is the coefficient that can be learned by training. It is considered LeakyReLU when S is a constant nonzero lower integer. 
Additionally, the proposed residual blocks can help reduce the computation cost, and the proposed MCDR can boost the performance reported in the ablation study.


\subsection{{Residual Compression and Filtering Network}}
\textbf{Residual Compression Network: } The $r_{t}$ is derived between $f_{t}$ and $\bar{f}_{t}$. Then it is encoded and decoded by the image compression network \cite{wu2020end}. 
The reconstructed version of $r_{t}$, denoted as $\hat{r}_{t}$, is obtained using the residual decoder. To compress $r_{t}$ efficiently, an autoencoder-type network based on convolution layers and Generalized Divisive Normalization (GDN) or Inverse GDN (IGDN) is used \cite{balle2018variational}. Similar to the motion estimation step, $r_{t}$ is transformed into $z_{t}$ after passing through the residual encoder. Subsequently, $z_{t}$ undergoes a quantization procedure, resulting in $\hat{z}{t}$. After the quantization, $\hat{z}{t}$ is decoded using the residual decoder, producing the reconstructed version $\bar{r}_{t}$.

\textbf{Residual Filtering Network: } 
After obtaining $\bar{r}_{t}$, which contains compression errors due to quantization in the residual encoder-decoder network, we introduce a residual filtering (RF) network to reduce these compression errors and improve the quality of the reconstructed frame. First, we warp $\bar{f}{t}$ and $\hat{f}{t-1}$ using a similar approach as in Equation \ref{eq1}. This results in a warped version of $\hat{f}{t-1}$, denoted as $\hat{f}{t-1}^{r}$. Then, we concatenate $\hat{f}{t-1}^{r}$ with $\bar{r}{t}$ and pass them through the RF network. The RF network consists of five convolution layers, each followed by a parametric rectified linear unit (PReLU), except for the final convolution layer. The convolution layer has three parameters: the size of the kernel, the output channel, and the stride, denoted as $\mathrm{k} \times \mathrm{n} \times \mathrm{s}$, respectively. Additionally, the RF network includes two upsampling layers and six residual blocks. The purpose of the upsampling layers is to minimize the computational requirements while preserving the input and output dimensions. The architecture of the residual blocks is the same as the residual block in the MCDR network. By applying the RF network, we obtain $\hat{r}{t}$, which represents the filtered residual frame and exhibits improved quality compared to $\bar{r}{t}$. The RF network effectively filters out compression errors, enhancing the visual quality of the reconstructed frame. In the ablation study, we further discuss the effectiveness of the RF network and provide visualizations of $\hat{r}_{t}$. This analysis helps evaluate the performance of the RF network in reducing compression errors and enhancing the overall quality of the reconstructed frames.

\subsection{Loss and Entropy Coding}
\textbf{Loss Function.} The goal of our training is to achieve video encoding with the smallest feasible number of bits while minimizing the distortion between $f_{t}$ and $\hat{f}_{t}$. We use a rate distortion, given by:
\begin{equation}
\alpha \mathrm{D}+\mathrm{R}=\alpha \mathrm{d}\left(f_{t}, \hat{f}{t}\right)+\left(\mathrm{N}\left(\hat{a}{t}\right)+\mathrm{N}\left(\hat{z}_{t}\right)\right)
\end{equation}
Here, $\mathrm{d}\left(f_{t}, \hat{f}_{t}\right)$ represents the distortion between $f_{t}$ and $\hat{f}_{t}$. The number of bits utilized for encoding is denoted by $\mathrm{N}(.)$. The bitstreams encode both the representations $\hat{a}{t}$ and $\hat{z}_{t}$, and a trade-off between the number of bits and distortion is controlled by $\alpha$. The quantization step is required before entropy coding, which involves the representations $a_{t}$ and $z_{t}$. However, differentiating the quantization process is not feasible for end-to-end training. To overcome this challenge, we adopt the approach proposed in \cite{wu2018compressed}. We substitute the quantization procedure with uniform noise, such that $\hat{a}{t}=a{t}+U$ and $\hat{z}{t}=z{t}+U$, where $U$ represents the uniform noise. This allows us to directly utilize rounding techniques in the prediction step, resulting in $\hat{a}{t}=round(a{t})$ and $\hat{z}{t}=round(z{t})$.


\textbf{Entropy Coding.} To improve the network, a bit rate estimate network is required. The estimation of the bit rate is crucial for accurately evaluating the entropy of the appropriate latent representation symbols. To achieve this, we employ the hyperprior entropy model, as introduced in \cite{minnen2018joint}. The hyperprior entropy model allows for precise estimation of the bit rate $\left(\mathrm{N}\left(\hat{a}{t}\right)\right.$ and $\left.\mathrm{N}\left(\hat{z}{t}\right)\right)$ of the quantized forms $\hat{a}{t}$ and $\hat{z}{t}$. By utilizing this model, we can accurately estimate the number of bits required to represent the latent representations of the motion vectors and residuals.



\section{Experiment}
\label{Dataset}
In the training step, we utilize the Vimeo-90K dataset \cite{xue2019video} that comprises 89,800 video clips, each with seven frames at a size of 448$\times$256 pixels. We use the following four datasets to compare our compression performance with state-of-the-art methods. \textbf{HEVC Test Sequences \cite{sullivan2012overview}:} It is the most common test sequence for assessing video compression performance. We use three classes including Class B ($1920\times1080$), Class C ($832\times480$), and Class D ($352\times288$) datasets. \textbf{MCL-JCV Dataset \cite{wang2016mcl}:} It consists of 24 video clips with a $1920\times1080$ resolution. This dataset is commonly used to judge the quality of the video. \textbf{Ultra Video Group (UVG) Dataset \cite{mercat2020uvg}:} It is a video dataset with a high frame rate ($120 fps$), in which the motion between adjacent frames is limited. According to the settings described in \cite{wu2018video, habibian2019video, lu2019dvc}, we conduct our studies of video sequences with $1920\times1080$ resolutions. \textbf{Video Trace Library (VTL) Dataset \cite{hu2021fvc}:} It comprises many raw YUV sequences (20 videos with $352\times288$ resolutions) utilized for low-level computer vision applications. 

\subsection{Implementation Details} 

During the training process, different $\alpha$ values are used for training. For the Mean Square Error (MSE) loss, $\alpha$ values of 64, 128, 256, and 512 are employed. For the Multi-Scale Structural Similarity Index (MS-SSIM) loss, $\alpha$ values of 16, 32, 64, and 128 are used. The training of the proposed model is conducted in two steps. In the first step, initial training is performed with a high bit rate, and the $\alpha$ value is set to 1024. This training lasts for 2,000,000 steps using the MSE loss. After the initial training, the pre-trained model is fine-tuned for an additional 500,000 iterations. This fine-tuning is done with $\alpha$ values of 64, 128, 256, and 512. 
During the fine-tuning phases, the learning rate for the newly added modules is initially set to $4e-5$ and is reduced to $1e-5$ for the final learning rate. The training is conducted with a batch size of 4. Overall, this training strategy allows the proposed model to be trained with different alpha values, fine-tuning the model to improve performance on both MSE and MS-SSIM criteria.

\begin{figure*}[htb]
    \centering
    \includegraphics[width=0.99\textwidth]{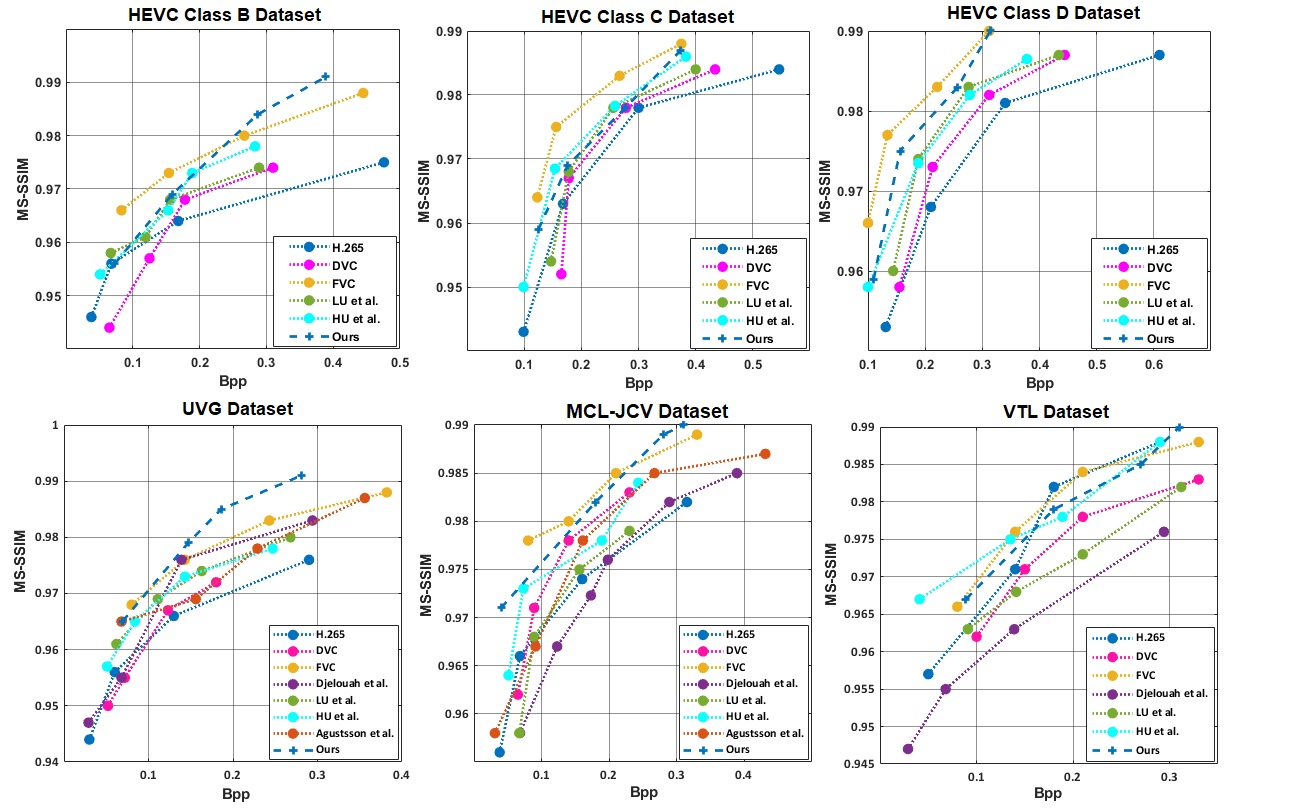}
    \caption{The MS-SSIM comparison of our method with the state-of-the-art learning-based methods, e.g., DVC \cite{lu2019dvc}, Djelouah et al. \cite{djelouah2019neural}, Agustsson et al. \cite{agustsson2020scale}, HU et al. \cite{hu2020improving}, LU et al. \cite{lu2020content}, FVC \cite{hu2021fvc}, and conventional methods, H.264/H.265 \cite{wiegand2003overview, sullivan2012overview} on HEVC Test Sequences (Class B, C, and D) \cite{sullivan2012overview}, UVG \cite{mercat2020uvg}, MCL-JCV \cite{wang2016mcl}, and VTL \cite{hu2021fvc} datasets.}
    \label{fig:f8}
\end{figure*}

\subsection{Evaluation} 

We utilize several metrics: Peak Signal-to-Noise Ratio (PSNR), Multi-Scale Structural Similarity Index (MS-SSIM), and Bjøntegaard delta bit rate (BDBR). The PSNR and MS-SSIM are computed by averaging the values across all reconstructed frames in each video sequence. These metrics provide insights into the quality of the reconstructed frames compared to the original frames. The BDBR metric measures the average number of bits required for motion and residual coding in each frame, represented as bits per pixel (Bpp). It quantifies the compression efficiency of the video codec. In our evaluation, learning-based approaches are compared to traditional codecs. While traditional codecs typically have a fixed Group of Pictures (GOP) configuration, our learning-based codec does not have such restrictions. We use a GOP size of 12 for datasets like UVG, MCL-JCV, and VTL, while for the HEVC Sequences, the GOP size is set to 10. This setup allows for a fair comparison between the learning-based video codec and conventional codecs, considering the impact of different GOP configurations on the rate-distortion performance.

\section{Experimental Results}
\label{Results}

\subsection{Computational Performance} 
Our model is implemented using PyTorch, taking advantage of CUDA support for efficient GPU acceleration. The experiments are conducted on a Windows 10 workstation with the following specifications: an Intel Core i7 CPU, 32GB of RAM, and a single NVIDIA GeForce RTX 2070 GPU with 8GB memory. The training process is computationally intensive and time-consuming. The initial training stage typically takes approximately seven days to complete. The second training stage, involving further fine-tuning of the model, lasts around five days. Finally, the fine-tuning step takes approximately one day to finish. These time estimates provide an overview of the computational requirements for training the model using the specified hardware setup. When encoding videos with a resolution of $352\times288$ pixels, the encoding speed of each iteration is 22.12fps (resp. 37fps), while Wu et al. \cite{wu2018video} and DVC \cite{lu2019dvc} are 29fps (resp. 38fps), and 24.5fps (resp. 41fps) on a single Titan 1080Ti GPU. We estimate that our proposed model's parameters are around 10.7M, especially for deep feature channels in autoencoder networks.

\subsection{Quantitative Performance}

\subsubsection{MS-SSIM Evaluation} Fig. \ref{fig:f8} demonstrates the rate-distortion curves of MS-SSIM for our method with state-of-the-art, e.g., conventional H.264/H.265 \cite{wiegand2003overview, sullivan2012overview},  deep learning-based methods DVC \cite{lu2019dvc}, Djelouah et al. \cite{djelouah2019neural}, Agustsson et al. \cite{agustsson2020scale}, HU et al. \cite{hu2020improving}, LU et al. \cite{lu2020content}, and FVC \cite{hu2021fvc} in all four datasets. Our method shows a promising MS-SSIM performance over state-of-the-art methods. In the HEVC class D dataset, our method received the same MS-SSIM result ($0.99$) as FVC \cite{hu2021fvc}, and for the HEVC class C dataset, it received the second-best performance of $0.987$ with around $0.38$ Bpp while FVC got about $0.988$ with the same Bpp. Thus, it generates higher-quality reconstructed frames. {Our model outperformed existing methods by a large margin in the MS-SSIM-based rate-distortion performance.}


\begin{figure*}[htb]
    \centering
    \includegraphics[width=0.99\textwidth]{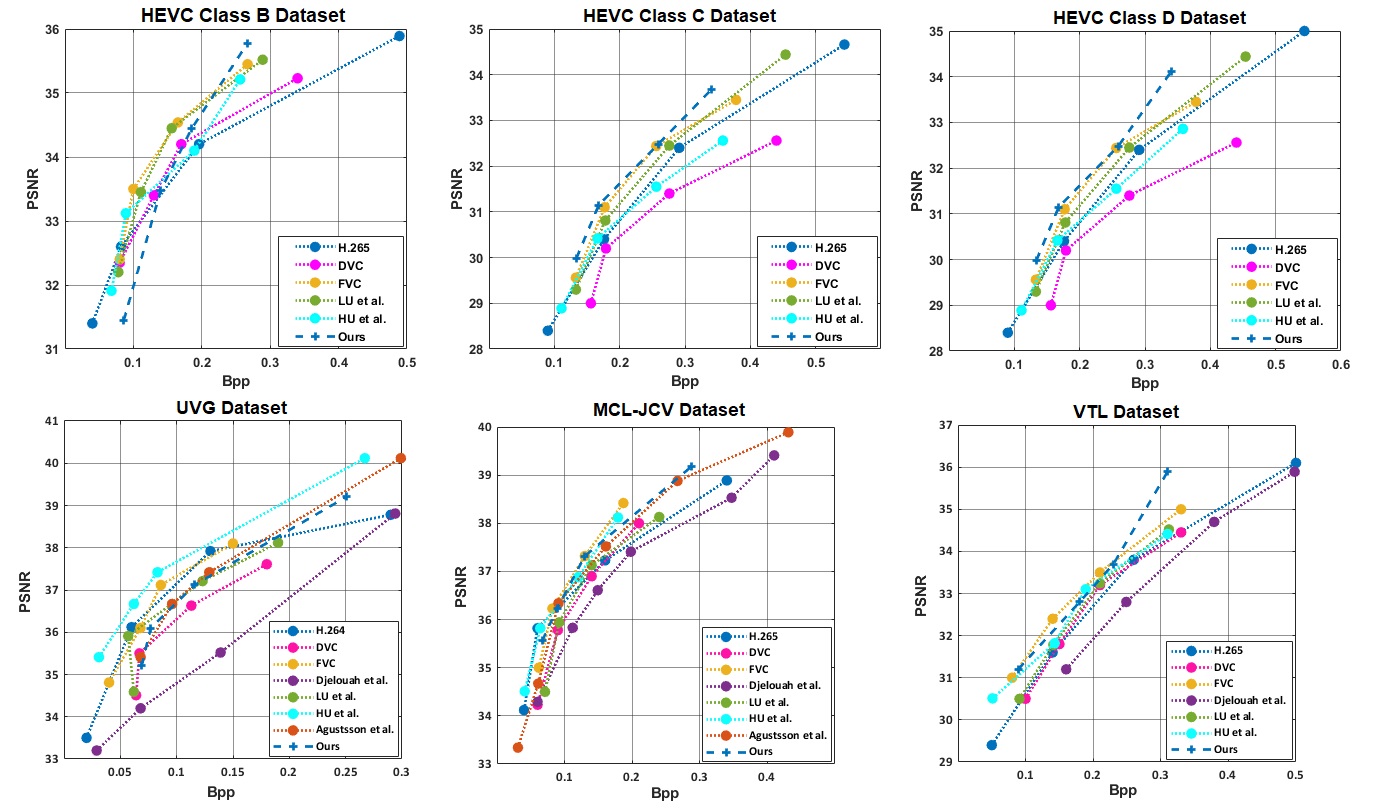}
    \caption{
    The PSNR comparison of our method with the state-of-the-art learning-based methods, e.g., DVC \cite{lu2019dvc}, Djelouah et al. \cite{djelouah2019neural}, Agustsson et al. \cite{agustsson2020scale}, HU et al. \cite{hu2020improving}, LU et al. \cite{lu2020content}, FVC \cite{hu2021fvc}, and conventional methods, H.264/H.265 \cite{wiegand2003overview, sullivan2012overview} on HEVC Test Sequences (Class B, C, and D) \cite{sullivan2012overview}, UVG \cite{mercat2020uvg}, MCL-JCV \cite{wang2016mcl}, and VTL \cite{hu2021fvc} datasets.}
    \label{fig:f9}
\end{figure*}

\begin{table}[htb]
\centering
\caption{Comparing the BDBR based on H.265. The bit-rate savings are shown by negative BDBR values}
\scalebox{0.62}{
\begin{tabular}{|c|c|c|c|c|c|c|}
\hline Dataset & DVC\cite{lu2019dvc} & Agustsson et al. \cite{agustsson2020scale} & LU et al.\cite{lu2020content} & HU et al.\cite{hu2020improving} & FVC\cite{hu2021fvc} & Ours \\
\hline UVG \cite{mercat2020uvg} & $6.35$ & $-8.64$ & $-8.27$ & $-15.34$ & $-27.71$ & $\mathbf{-30.89}$ \\
\hline MCL-JCV \cite{wang2016mcl} & $15.94$ & $-3.62$ & $3.64$ & $-12.67$ & $\mathbf{-23.48}$ & $-22.14$ \\
\hline VTL \cite{hu2021fvc} & $-8.83$ & $-$ & $-19.78$ & $-25.11$ & $-27.10$ & $\mathbf{-29.21}$ \\
\hline HEVC Class B \cite{sullivan2012overview} & $5.86$ & $-$ & $-16.80$ & $-13.73$ & $-24.54$ & $\mathbf{-27.51}$ \\
\hline HEVC Class C \cite{sullivan2012overview}  & $18.65$ & $-$ & $-5.65$ & $2.12$ & $-14.18$ & $\mathbf{-16.78}$ \\
\hline HEVC Class D \cite{sullivan2012overview} & $16.10$ & $-$ & $-7.89$ & $-2.34$ & $\mathbf{-18.39}$ & $-16.11$ \\
\hline
\end{tabular}}
\label{tab:my_label}
\end{table}

\subsubsection{PSNR Evaluation} From the rate-distortion curves of PSNR in Fig. \ref{fig:f9}, it can be shown that our method outperforms both the baseline methods, DVC \cite{lu2019dvc} and FVC \cite{hu2021fvc} in all tested datasets. On HEVC class B datasets \cite{sullivan2012overview}, the PSNR/Bpp of our approach is 38.8/0.278, while the PSNR/Bpp is about 35.2/0.333 and 35.5/0.279 for both baseline methods. For the VTL dataset \cite{hu2021fvc}, our method yields around 39/0.321 (PSNR/Bpp), while the DVC and FVC methods achieve approximately 34/0.344 and 35/0.333 (PSNR/Bpp), respectively. The proposed method surpasses the most popular conventional techniques H.264 (for UVG dataset \cite{mercat2020uvg}, around PSNR, 39.2 vs. 38.8 and Bpp, 0.25 vs. 0.28) and H.265 (for MCL-JCV dataset \cite{wang2016mcl}, around PSNR, 39.2 vs. 39 and Bpp, 0.28 vs. 0.34), respectively. {Furthermore, it exceeds LU et al. \cite{lu2020content}. on all datasets except HEVC Class C and D \cite{sullivan2012overview} and outperforms HU et al. \cite{hu2020improving} on every dataset except the UVG \cite{mercat2020uvg}.} 

\subsubsection{BDBR Evaluation} In TABLE \ref{tab:my_label}, we present the BDBR \cite{bjontegaard2001calculation} findings of our method with state-of-the-art methods on H.265. The bit saving rate of our method is significantly higher, which is 30.89\% for the UVG dataset, whereas the recent approach FVC \cite{hu2021fvc} is 27.71\% and for Agustsson et al. \cite{agustsson2020scale}, LU et al. \cite{lu2020content}, and HU et al. \cite{hu2020improving} are 8.64\%, 8.27\%, and 15.34\%, respectively. For VTL, HEVC class B, and C datasets, our method also performed considerably better in terms of bit savings rate, with 29.21\%, 27.51, and 16.78\%, respectively. However, the bit savings rate is slightly lower than FVC (23.48 vs. 22.14 and 18.39 vs. 16.11) in VTL and HEVC class D datasets. Furthermore, our technique outperforms the baseline method DVC on all datasets.

\begin{figure*}[htb]
    \centering
    \includegraphics[width=0.425\textwidth]{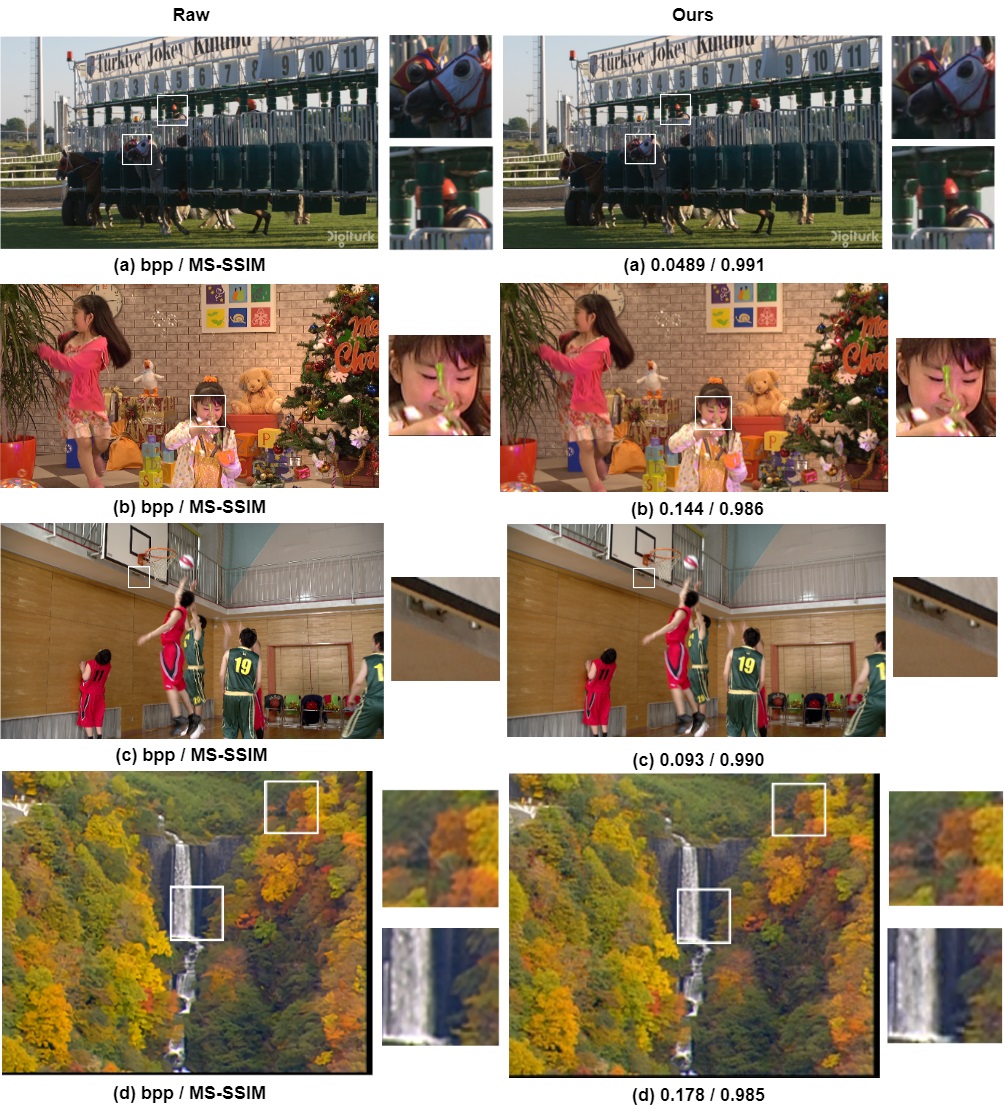} 
    \hspace{0.2cm}
    \includegraphics[width=0.51\textwidth]{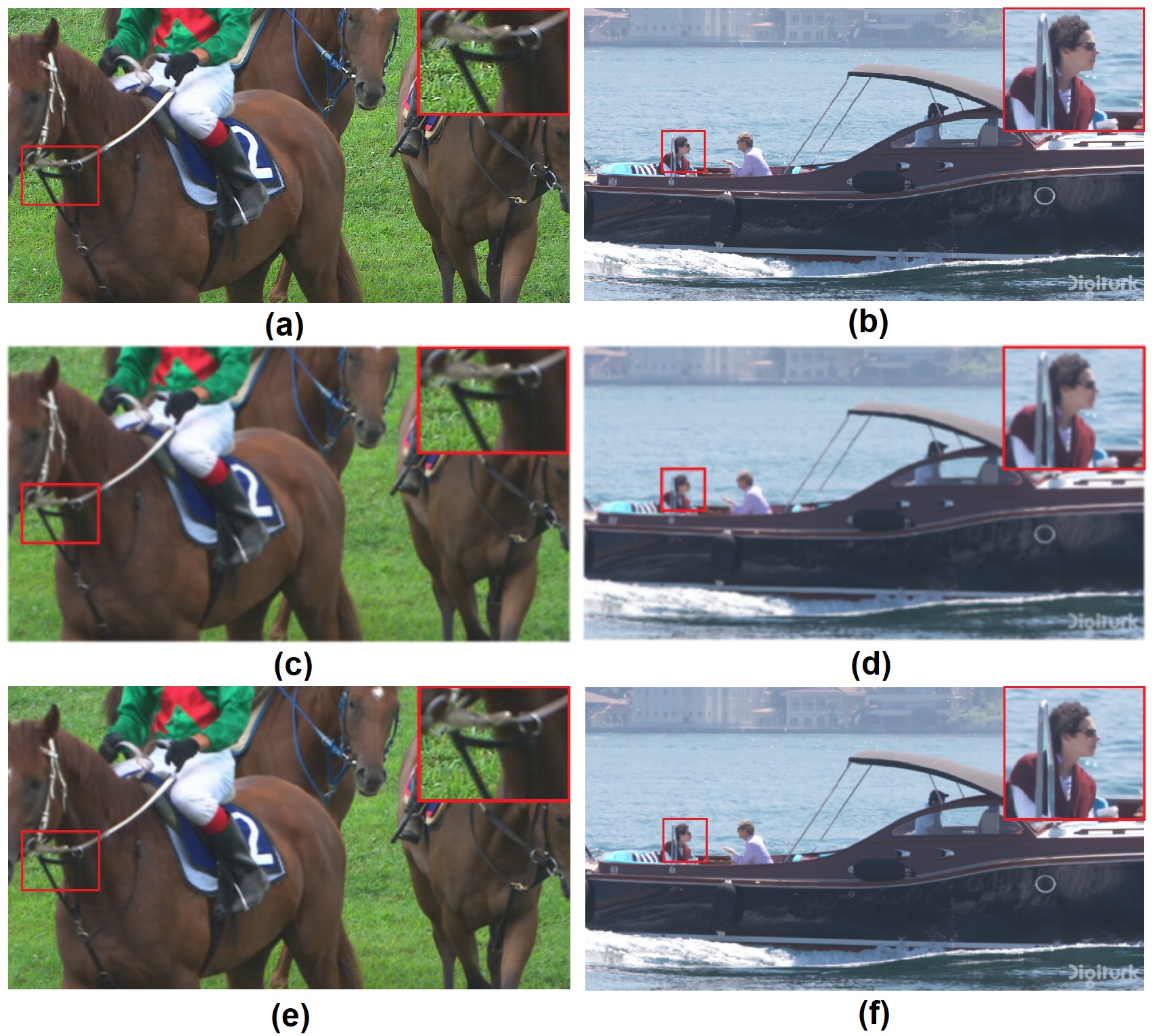}    
    \caption{Left: Qualitative comparison between the original and our reconstruction frames on the UVG (a), HEVC (b), MCL-JCV (c), and VTL (d) dataset. Right: The predicted video frames without, and with considering our MVF module. Original frames from  HEVC and UVG dataset (a, b), predicted frames without (c, d), and with (e, f) considering MVF module in our method.}
    \label{fig:f10}
\end{figure*}


\begin{figure}[htb]
    \centering
    \includegraphics[width=0.5\textwidth]{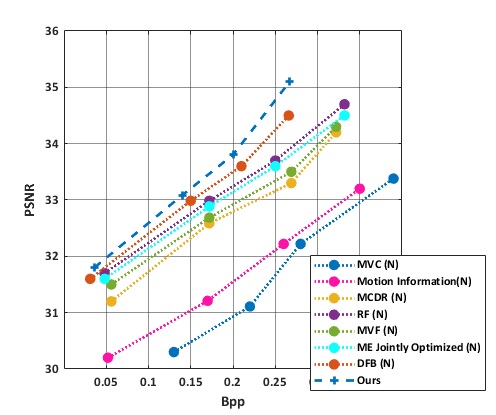}
    \caption{Ablation study of different modules on the HEVC Class B dataset \cite{sullivan2012overview}. The blue dashed line represents our result. {The dotted lines represent the results by removing the motion vector compression network (MVC(N)), motion information network, motion compensation network (MCDR(N)), residual filtering (RF(N)) network, motion vector filtering (MVF(N)) network, the motion estimation (ME) module,} and a decoded frame buffer (DFB(N)).}
    \label{fig:f11}
\end{figure}


\subsection{Qualitative Result} 

In Fig. \ref{fig:f10} (left), we present a qualitative comparison between the original and reconstructed frames. The bit rates represent the average of all frames in each video. Based on the MS-SSIM and {Bpp}, our reconstructed frames show great compression by using a lower bit rate. The reconstructed frame in the VTL dataset achieves a 0.985 MS-SSIM with 0.178 Bpp. Fig. \ref{fig:f10} (right) displays the qualitative results of the predicted frames without considering the {MVF} module. The predicted results without the {MVF} module exhibit more artifacts with noise. On the other hand, the results of our method (considering the {MVF} module) are considerably smoother, as seen in the grasses (e) and the boat region (f), with fewer artifacts and improved visualization. Moreover, the PSNR value for (c) is 32.44 dB, and for (d) is 32.15 dB, while it is significantly higher for (e) at 34.14 dB and (f) at 34.47 dB. Thus, our MCDR network produces more accurate results by incorporating the {MVF} model.

\section{Ablation Study}
\label{Ablation Study}

\noindent \textbf{Effectiveness of Contributed Modules:} Fig. \ref{fig:f11} shows the ablation study performance on different modules on the HEVC Class B dataset. For example, we consider each frame separately without considering any motion estimation approaches to illustrate the efficacy of motion information for video compression (see the pink dotted curve represented by Motion Information (N)). As a result, the PSNR performance is reduced by roughly 1.9 dB compared to ours (35.1 to 33.2 dB). The encoder and decoder compression network are removed (see the dotted blue line curve represented by MVC (N)). We notice that the system's performance dropped 1.3 dB compared to our baseline technique with high bits per pixel (around 0.39 vs. 0.28). We employ a motion compensation network with residual connection and deep rectifiers (PReLU) to build the predicted frame. To demonstrate the efficacy of the motion compensation network, we perform an experiment removing the MCDR network, shown by the yellow dotted line in Fig. \ref{fig:f11} (MCDR (N)). It is noticeable that the PSNR is reduced by 0.9 dB at the 0.322 Bpp level. Additional experiments are conducted on the proposed {MVF and RF} by removing them to observe their usefulness (see the dotted green and purple line curves). The system's performance dropped by 0.8 dB (35.1 vs. 34.3 for removing {MVF}) and 0.4 dB (35.1 vs. 34.7 for removing {RF}). To illustrate the efficiency of the iterative buffer, we draw it during the experiment (see the dotted orange line curve denoted by DFB (N)). As a result, the PSNR value is reduced by 0.6dB (35.1 vs. 34.5). Our method utilizes the end-to-end training strategy to improve the motion estimate module across the entire network. When the joint training is eliminated throughout the experiment, our method's performance drops by 0.6dB (35.1 vs. 34.5) with a larger Bpp (0.332). Thus, each module is crucial for our proposed method. 


\begin{table}[htb]
\centering
\caption{Effectiveness of output feature channels in autoencoder network on inference time and other metrics.}
\scalebox{0.8}{
\begin{tabular}{|c|c|c|c|c|c|c|c|}
\hline
Feature Channel (FC) & 32 & 64 & 128 & 256 & 512 & 1024 & 2048 \\ \hline
PSNR (dB) & 35.56 & 36.32 & 37.13 & 38.11 & 38.39 & 38.40 & 38.41 \\ \hline
MS-SSIM & 0.972 & 0.979 & 0.984 & 0.988 & 0.989 & 0.9895 & 0.9890 \\ \hline
Inference time (ms) & 24 & 26 & 29 & 31 & 33 & 44 & 53 \\ \hline
Num. of parameters (M) & 0.89 & 1.14 & 2.22 & 3.21 & 4.39 & 7.85 & 9.72 \\ \hline
\end{tabular}}
\label{tab:channel}
\end{table}

\noindent \textbf{Effectiveness of Output Feature Channels:}
The inference time for a single frame should not exceed 33 milliseconds (ms) to achieve optimal quality. We measured the end-to-end inference time per frame for different output feature channels (FC), as presented in TABLE \ref{tab:channel}. It can be observed that increasing the number of output feature channels (FC) leads to improvements in PSNR, MS-SSIM, inference time, and parameters. However, after reaching an FC of 512, further increases in FC result in significant increases in inference time and the number of parameters, while only slightly improving PSNR and MS-SSIM. For instance, when increasing the FC from 512 to 2048, the number of parameters increases by approximately 55\% (9.72 million vs. 4.39 million), while the inference time rises by approximately 38\% (53 ms vs. 33 ms). 


\section{Conclusion}
\label{Conclusion}

In this paper, we presented an end-to-end video compression method that effectively reduces temporal redundancy and rate-distortion tradeoff. The framework consists of several key components. Firstly, a deep recurrent feature pyramid-based network was introduced to accurately capture motion information from optical flow in the motion estimation module. This allows for precise motion estimation, which was essential for efficient video compression. Next, a residual learning-based autoencoder-style network was employed to encode the optical flow using powerful nonlinear transformations, such as Generalized Divisive Normalization (GDN). This enables deeper learning and captures more nuanced features in the optical flow. To further improve compression performance, two filtering networks, MVF and RF, were introduced. These networks, equipped with deep rectifier PReLU, reduce compression errors in the encoder-decoder network and enhance the quality of predicted and residual frames. The inclusion of an update buffer ensured accurate reference frames throughout the training phase, contributing to improved compression results. 
In future work, we plan to explore the development of an entropy model to further enhance compression efficiency, aiming for even more effective and efficient video compression and overcoming the limitations mentioned in the earlier subsection.

\section*{Acknowledgement} This work is partially supported by the Scientific and Technological Research Council of Turkey (TUBITAK) under the 2232 Outstanding Researchers program, Project No. 118C301. 

\bibliographystyle{IEEEtran}
{\small
\bibliography{refs}}

\end{document}